\documentclass[prc,twocolumn,showpacs,preprintnumbers,superscriptaddress,
               amsmath,amssymb]{revtex4}
\usepackage{graphicx}          
\usepackage{dcolumn}           
\usepackage{bm}                

\begin{document}

\preprint{\fbox{\sc version of \today}}

\title{Tetrahedral Symmetry in Ground- and Low-Lying States of Exotic 
       $\boldsymbol{A\sim 110}$ Nuclei}
       
\author{ N. Schunck}
\affiliation{
            {\it Department of Physics 
                 University of Surrey,
                 GU2 7XH Guildford, Surrey, UK}\\
	    }
\affiliation{
            {\it Institut de Recherches Subatomiques 
                 IN$_2$P$_3$-CNRS/Universit\'e Louis Pasteur \\
                 F-67037 Strasbourg Cedex 2, France}\\
            }
\author{J.~Dudek}
\affiliation{
            {\it Institut de Recherches Subatomiques 
                 IN$_2$P$_3$-CNRS/Universit\'e Louis Pasteur \\
                 F-67037 Strasbourg Cedex 2, France}\\
            }
\author{A.~G\'o\'zd\'z}
\affiliation{
            {\it Katedra Fizyki Teoretycznej, 
                 Uniwersytet Marii Curie-Sk\l{}odowskiej,}
                 PL-20031 Lublin, Poland
            }
\author{P.H.~Regan}
\affiliation{
            {\it Department of Physics 
                 University of Surrey,
                 GU2 7XH Guildford, Surrey, UK}\\
	    }

\date{\today}


\begin{abstract}             
       Recent calculations predict a possible existence  of nuclei with
       tetrahedral symmetry: more precisely, the mean-field hamiltonians of
       such nuclei are symmetric with respect to double point-group $T_d^D$. In
       this paper, we focus on the neutron-rich  Zirconium isotopes as an
       example and present realistic mean-field  calculations which predict
       {\em tetrahedral ground-state}  configurations in $^{108,110}$Zr and
       low-lying excited states of  tetrahedral symmetry in a number of $N \leq
       66$ isotopes. The  motivations for focusing on these nuclei together with
       a discussion of the possible experimental signatures of tetrahedral
       symmetry are  also presented.
\end{abstract} 


\maketitle

%
%

A highly relevant field of investigation of nuclei with unusual combinations of
the proton and neutron numbers is the search for manifestations of exotic
symmetries. In our previous work we reported \cite{JDu02}, that of
all the various possible non-spherical shape symmetries, tetrahedral and
octahedral ones may lead to the largest single-particle shell gaps. Indeed,
these can be comparable to, and sometimes even larger than the well-established 
spherical shell gaps. These properties, far from being restricted to nuclei 
within a given range of mass or isospin, should in fact be present throughout 
the nuclear chart. The underlying considerations are based on a very general 
analysis of the point groups of symmetries of the nuclear mean-fields. They 
have been confirmed independently via the macroscopic-microscopic method of 
Strutinsky \cite{JDu02} and the self-consistent Hartree-Fock techniques 
\cite{TYM98}.

The purpose of this Rapid Communication is to support  the search for
tetrahedral nuclei by providing realistic calculations in some  selected study
cases.

The neutron-rich Zirconium (Z=40) isotopes provide such an appropriate 
case. Our previous work predicted that, among others, nucleon numbers $Z=N=40$ 
and $Z=N=70$ correspond to large shell gaps for tetrahedral shapes and thus 
can be thought of as `tetrahedral magic numbers' \cite{JDu02,JDu02a}. Although 
the yrast line of the $N=Z=40$ system $^{80}$Zr has been experimentally 
studied above spins of 10$\hbar$ \cite{SMF87}, the detailed information
concerning in particular the electro-magnetic transitions (see below)
which is required for probing the tetrahedral symmetry has not
yet been achieved. On the neutron-rich side, Zirconium isotopes up to the
$N=70$ system  $^{110}$Zr have been synthesised using the projectile fission
technique at GSI, \cite{MBe94}. The production of such nuclei opens up the
possibility of studying their properties via isomer spectroscopy following
projectile fragmentation (e.g. \cite{ZPo00}) and projectile-fission (e.g.
\cite{MNM01}) or beta-delayed gamma-ray spectroscopy (e.g. \cite{PFM03}). To
date however, such studies have not yet been performed in the very heavy
Zirconium isotopes and the most neutron rich $Z=40$ isotope where information
on excited states is available is $^{104}$Zr$_{64}$, which was studied as a
fragment from spontaneous fission of Cm sources in \cite{JHH97}.

In this article we suggest that the heavy Zr isotopes are amongst the best
candidates to test the hypothesis of nuclear tetrahedral symmetry: 
$^{110,112}$Zr are indeed predicted by Hartree-Fock calculations to have 
$T_d^D$-symmetric {\it ground-state configurations}, and both 
non-self-consistent approaches and Hartree-Fock calculations agree on the 
prediction of low-lying excited $T_d^D$-symmetric minima for a chain of
isotopes near $^{110}$Zr.  As outlined above, these nuclei are viable for
future experimental study. 

In the basis of spherical harmonics, the distance of a point on the nuclear 
surface relative to the centre of the nucleus can be expressed by
\begin{equation}
       R(\theta,\varphi) 
       = 
       r_{0} \, A^{1/3} \, c( \alpha )
       \bigg[ 
             1 + \sum_{\lambda,\mu} \alpha_{\lambda\mu} 
                                         Y_{\lambda\mu}(\theta,\varphi) 
       \bigg],
                                                                  \label{eqn01}
\end{equation}
where $\alpha_{\lambda\mu}$ refer to multipole deformations. In
Eq.~(\ref{eqn01}), $\alpha \equiv \{ \alpha_{\lambda \, \mu}\}$  represents 
the complete set of deformations and $c(\alpha)$ accounts for nuclear volume 
conservation. Several ensembles of deformation parameters 
$\alpha \equiv \{ \alpha_{\lambda \, \mu}\}$ can generate a 
geometry characteristic of the group $T_{d}^{D}$. The simplest case occurs 
when all deformations except $\alpha_{32}$ are equal to 0. This simplest 
form of the tetrahedral deformation is also likely to be favoured in nature 
since other realizations of the tetrahedral symmetry involve {\em much higher} 
order of multipoles, $\lambda \geq 7$, cf. Ref.~\cite{JDu02a}.

The group $T_{d}^{D}$ is characterized by two non-equivalent two-dimensional 
irreps, and one {\em four-dimensional} one. This is an unique feature in
nature, since, apart  from the spherical symmetry, only the groups $T_{d}^{D}$
and $O_{h}^{D}$ (the  latter, octahedral symmetry is discussed in
Ref.~\cite{JDu02a}) produce  single-particle degeneracies higher than 2, cf.
Refs.~\cite{JFC84,GFK63}. This  high degeneracy pattern, together with a
relatively low number of irreps  favours the appearance of large shell gaps. In
our previous work \cite{JDu02},  we reported that such gaps are predicted in a
number of different mass regions,  and presented various examples of potential
tetrahedral minima.

\begin{figure}[h]
\vspace*{-0.15truecm}
\includegraphics[height=8.5cm,width=5.7cm,angle=-90]{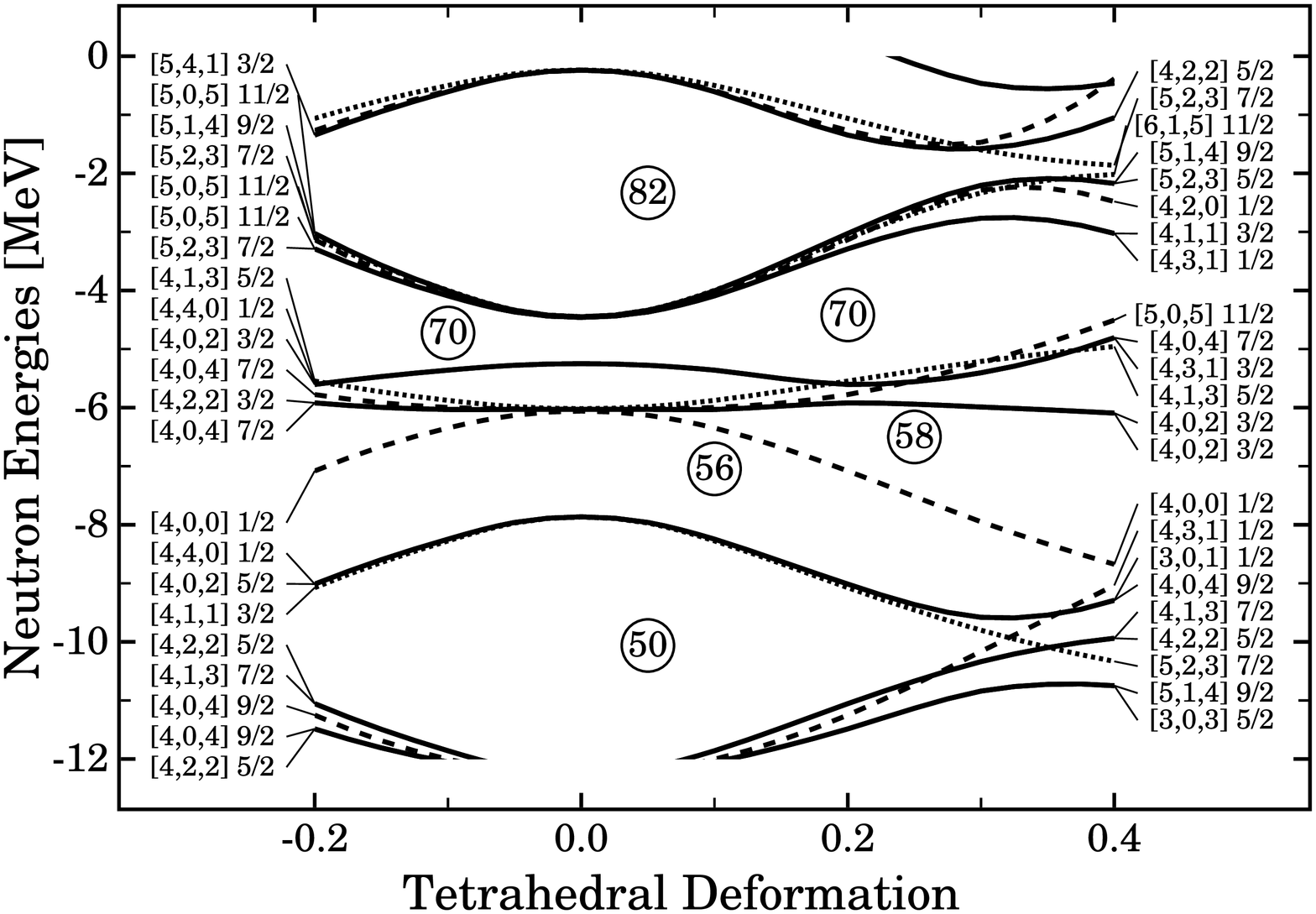}

\vspace*{-0.28truecm}

\includegraphics[height=8.5cm,width=5.7cm,angle=-90]{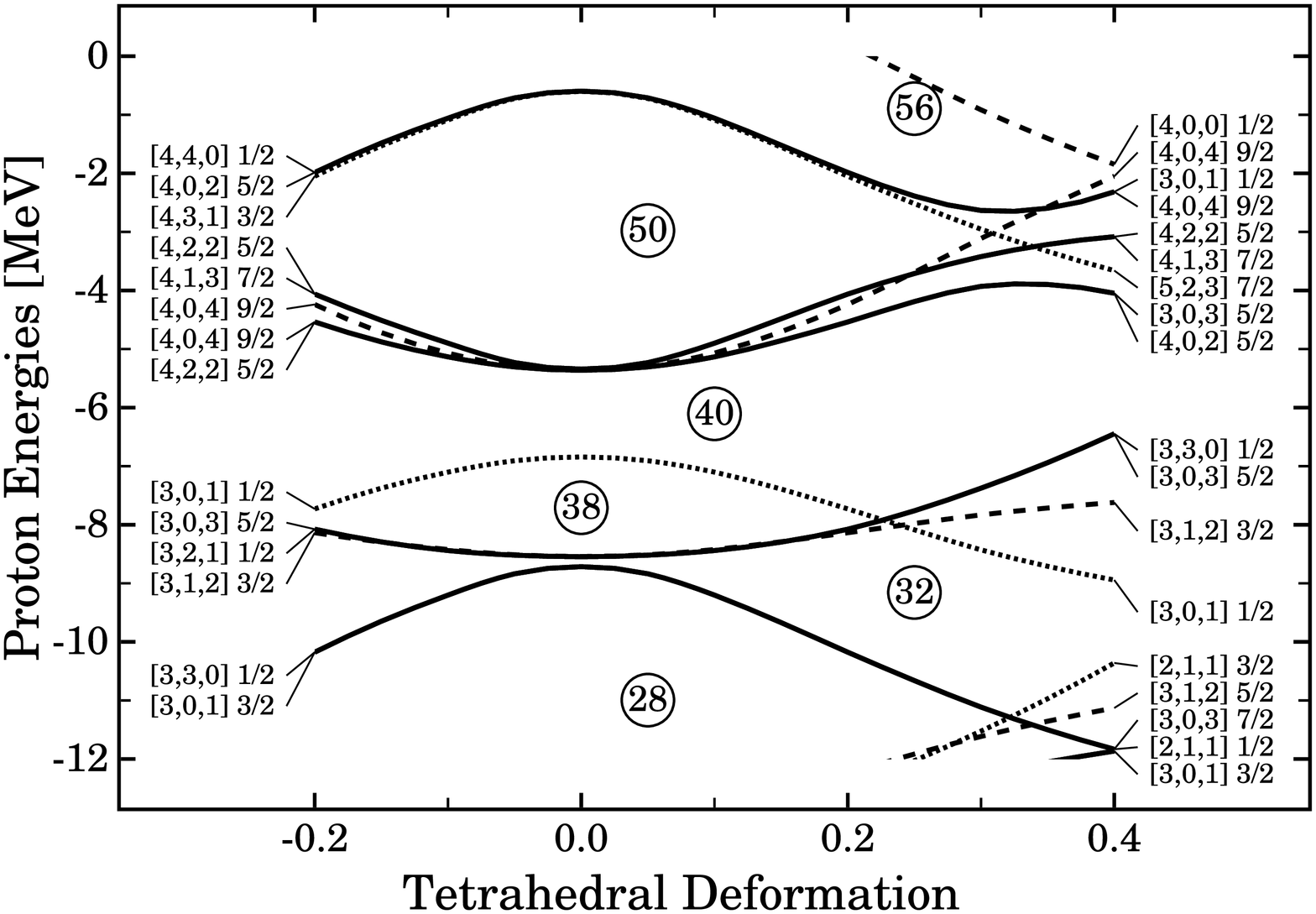}
\caption{                  
         Single-particle neutron (top) and proton (bottom) 
	 energies as a function of the lowest-order tetrahedral deformation 
	 $(=\alpha_{32})$ for nuclei  around $^{110}$Zr$_{70}$. The Nilsson 
	 labels correspond to the leading basis wave-function in the expansion 
	 of the single-particle wave function. The eigenenergies belonging to 
	 the four-dimensional irreps (thus forming 4-fold degenerate 
	 multiplets) are marked with the full lines and double Nilsson labels. 
	 The levels belonging to the two non-equivalent two-dimensional 
	 irreps are marked with dashed lines.
         }
                                                                  \label{fig01}
\vspace*{-0.2truecm}
\end{figure}

In the current work, realistic nuclear structure calculations are performed
within the standard Woods-Saxon mean-field approach  with the universal
parameterization. Figure \ref{fig01} shows the single-particle energies for
exotic nuclei with  $A \sim 110$ as functions of the lowest order tetrahedral
deformation obtained from this hamiltonian. The most striking result in 
Fig.~\ref{fig01} is the presence of a large tetrahedral shell gap at $Z=40$, 
comparable to the spherical $Z=50$ gap. Indeed this gap is {\em larger} by 
$\sim 3$ MeV than the spherical one at $Z=40$. 

To calculate the total energy of the nucleus, we employ the 
microscopic-macroscopic method whose realization closely follows that of
Refs.~\cite{JDu87} and \cite{TRW95}. In this approach, the total energy of the
nucleus is given by:
\begin{equation}
       E = E_{macro}+ E_{micro}\, ;
       \;        
       E_{micro} = \delta E_{shell} + \delta E_{pair}, 
                                                                  \label{Strut}
\end{equation}
where $E_{macro}$ is the liquid drop energy, here calculated as in
Ref.~\cite{KP002}; the latter realization includes in particular the nuclear
surface-curvature contributions. 

The quantum correction $\delta E_{shell}$, accounts for the nuclear shell
structure. It is calculated according to the standard Strutinsky prescription
\cite{VMS67}, and supplemented with the pairing correlation energy, $\delta
E_{pair} \equiv E_{pair} - E_{pair=0}$, obtained within the
pairing-selfconsistent Hartree-Fock-Bogolyubov (HFB) formalism as in
Ref.~\cite{MdV83}. The $E_{pair}$ term refers to the energy of the system in 
the presence of pairing, calculated by using the Particle Number Projection 
(PNP) technique (before variation), and $E_{pair=0}$ is the analogous term 
obtained by setting the pairing correlations to zero. The average pairing 
strength constants have been obtained in the usual way by fitting to the known 
nuclear masses of nuclei in the studied region.

The deformation space used includes {\it a priori} all possible
$\alpha_{\lambda\mu}$ up to $\lambda = 12$, with $-\lambda \leq \mu \leq
+\lambda$. However, the number of multipoles effectively used was significantly
reduced since the effect of high-$\lambda$ was found to be negligible (by a 
direct verification).

\begin{figure}[h]
\vspace{-0.25truecm}
\includegraphics[height=8.5cm,width=5.3cm,angle=-90]{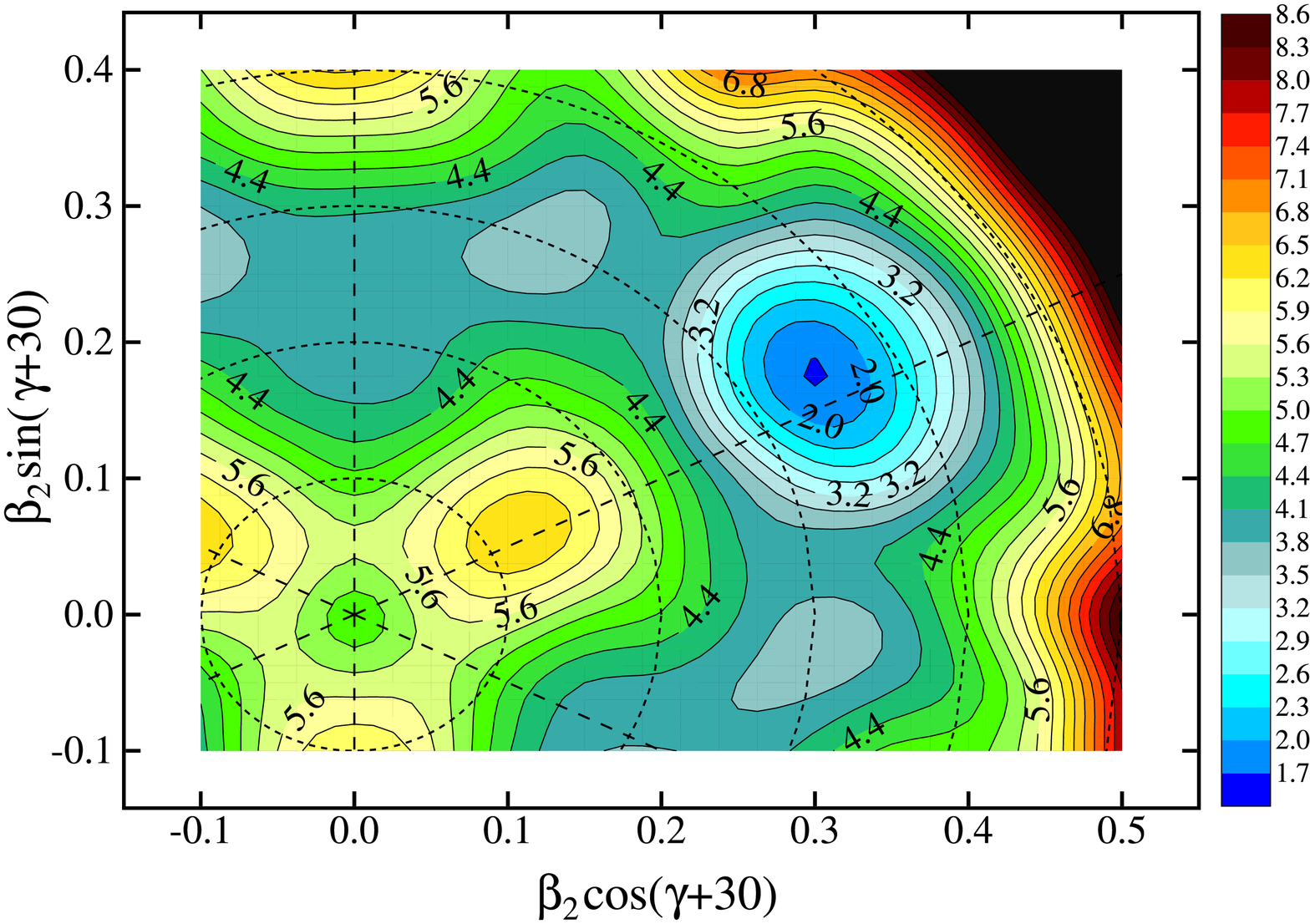}

\vspace*{-0.45truecm}

\includegraphics[height=8.5cm,width=5.3cm,angle=-90]{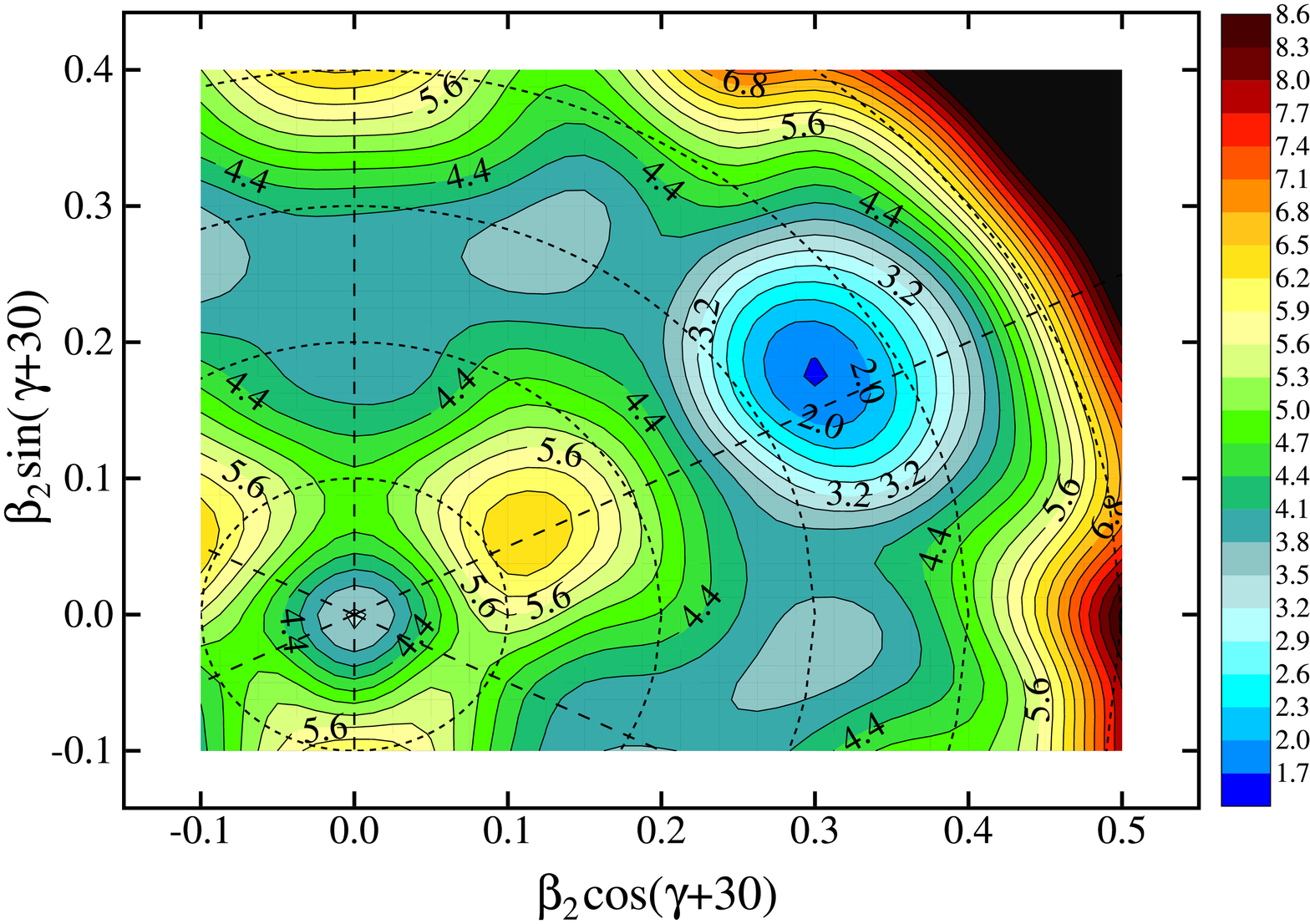}
\caption{
        (Color online) Ground-state total energy surfaces in $^{110}$Zr as a 
	function of the quadrupole $(\beta,\gamma)$ deformations. Top: 
	minimisation over $\alpha_{40}$ at each $(\beta,\gamma)$-point; 
	bottom: similar to the above but minimized over tetrahedral 
	deformation $\alpha_{32}$. Note that the tetrahedrally-symmetric 
	minimum is lower by more than 1 MeV than the spherical minimum.
        }
                                                                  \label{fig02}
\vspace*{-0.2truecm}
\end{figure}

The potential energy landscapes of Fig.~\ref{fig02}, top, confirm the earlier
calculations by \cite{TRW95,FRX02,skalski,PMo95}, all of which predict that at 
its ground-state, the $^{110}$Zr nucleus has a significant prolate deformation 
with an axial hexadecapole deformation $\alpha_{40}$ of about 0.10. 
The calculation for the top frame of Fig.~\ref{fig02} has been performed 
according to the general procedure discussed above, but the deformation space 
included only the quadrupole $(\beta_2, \gamma)$ and axial hexadecapole 
$\alpha_{40}$ deformation (in the following we employ the standard Hill-Wheeler 
parameterization for quadrupole deformations). Beside the prolate ground-state, 
we note the presence of a very flat oblate pattern together with a small 
spherical minimum. However, the inclusion of the tetrahedral degree of freedom 
has a strong impact on this energy landscape. While the prolate minimum 
remains, the minimum at $\beta_2 = 0$ is lowered in energy by more than 1 MeV 
and corresponds to a tetrahedral configuration with $\alpha_{32} \sim 0.14$. 
In the calculations this minimum appears as pure i.e. all other deformations
turn out to be zero as the result of the minimization. 

The often employed technique of calculations using a mesh of deformation points
(as in Fig.~\ref{fig02}) has the drawback of taking into account only a very
partial projection (usually less than 4 deformations) out of the full 
deformation space. In order to remove this limitation, we have adopted a 
strategy based on a dynamical minimization of the total energy: for fixed 
values of a given deformation parameter (in our case the elongation along the 
$z$-axis, $\beta_{2}$), the total energy was minimized with 
respect to many other deformation variables using the standard Variable Metric 
Method described in \cite{WHP94}. In our calculations the deformation space is
composed of the entire set of octupole $\{ \alpha_{3\mu}; \; \mu=0,1,2,3 \}$ 
and hexadecapole $\{ \alpha_{4\mu}; \; \mu=0,1, \ldots,4 \}$ degrees of
freedom, in addition to the $\gamma$ angle, i.e. 11 deformation parameters in
total. As mentioned above, the effect of higher order multipoles has been
investigated and is neglegible in the present context.

\begin{figure}[h]
\vspace{-0.15truecm}
\begin{center}
\includegraphics[height=6.0cm,width=8.0cm]{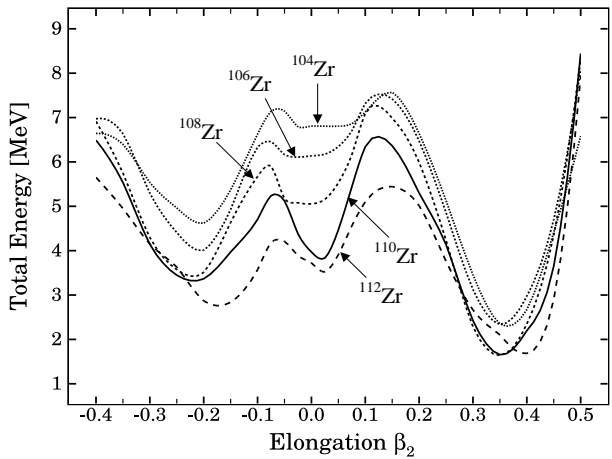}
\vspace*{-0.6truecm}
\end{center}
\caption{  
         Total energy as function of quadrupole deformation in a 
	 chain of Zr isotopes. Negative values of $\beta_{2}$ correspond
	 to oblate shapes. Each $\beta_{2} \sim 0 $ minimum correspond to a
	 tetrahedral configuration. (The pairing correlations were included 
	 and treated with the help of the PNP technique).
         }
                                                                  \label{fig03}
\vspace*{-0.2truecm}
\end{figure}

Figure~\ref{fig03} shows the minimized-energy cross-sections
for the ground-states in several isotopes of Zirconium as function of the
quadrupole deformation $\beta_2$. For each isotope three
minima appear: one at an elongation of $\beta_2 \sim 0.35-0.40$, another one
around $\beta_2 \sim -0.25$, and a third one at $\beta_2 \sim 0$. The latter 
corresponds to a finite tetrahedral deformation of about 
$\alpha_{32} \sim 0.14$. The curves plotted in Fig.~\ref{fig03} may 
provide an estimate of the height of the barriers between the tetrahedral 
minima and the other deformed minima. In the case of $^{110}$Zr, this barrier 
amounts to up to $\sim$ 1.5 MeV with respect to the oblate minimum.

\begin{table}[ht]
\vspace*{-0.1truecm}
\begin{center}
\caption{HFB energies (in MeV) for various energy minima in heavy Zr isotopes, 
         relative to the energy of the tetrahedral minimum. Calculations were
	 done with the SLy4 force \cite{Cha97}.}
\vspace*{+0.1truecm}
\label{table1}
\begin{tabular}{ccccccc}
\hspace*{0.2cm}Nucleus\hspace*{0.2cm} & \hspace*{0.2cm}$^{104}$Zr\hspace*{0.25cm} & \hspace*{0.2cm}$^{106}$Zr\hspace*{0.2cm} 
& \hspace*{0.2cm}$^{108}$Zr\hspace*{0.2cm} & \hspace*{0.2cm}$^{110}$Zr\hspace*{0.2cm} & \hspace*{0.15cm}$^{112}$Zr\\ \hline\hline
\end{tabular}
\begin{tabular}{cdddddd}
Tetrahedral &  +0.00\hphantom{0} & +0.00\hphantom{0} & +0.00\hphantom{0} & +0.00\hphantom{0} & +0.00    \\
Spherical   &  +0.22\hphantom{0} & +0.29\hphantom{0} & +0.39\hphantom{0} & +0.43\hphantom{0} & +0.03    \\
Oblate      &  -1.57\hphantom{0} & -1.52\hphantom{0} & -1.10\hphantom{0} & +0.07\hphantom{0} & +0.30    \\
Prolate     &  -2.07\hphantom{0} & -1.76\hphantom{0} & -0.68\hphantom{0} & +0.27\hphantom{0} & +1.01    \\
\end{tabular}
\end{center}
\vspace*{-0.4truecm}
\end{table}
Self-consistent Hartree-Fock-Bogliubov calculations have also been performed 
for these nuclei, using the SLy4 force \cite{Cha97} and the {\sf HFODD} code
(version 2.07) of \cite{hfodd}, and are reported in Table \ref{table1}. The 
same energy profiles as shown in Fig.~\ref{fig03} are predicted, with the 
{\it noticeable exception} that the tetrahedral minimum in $^{110}$Zr and 
$^{112}$Zr lies {\it lower} in energy than the now secondary prolate minimum. 
Both Fig.~\ref{fig03} and Table \ref{table1} show that the influence of the 
shell effects leading to the tetrahedral symmetry diminishes when going away 
from the 'tetrahedral magic gap' $N=70$. For the lighter isotopes like 
$^{104}$Zr and $^{106}$Zr, the tetrahedral minimum lies very high in energy 
and is extremely shallow.

An important aspect associated with the preceding discussion is to find a
possibly unambigous experimental signature of the discussed symmetry.
Here we would like to construct  a criterion based on the rotational properties
of a nucleus with the tetrahedral symmetry. A discussion of other possible 
experimental signatures can  be found in \cite{JDu02a}. We use a
generalisation of the rotor hamiltonian often used in molecular physics, cf.
e.g. \cite{Har93}. We define
\begin{equation}
      \hat{\mathcal{H}} 
      = 
      \hat{I}^{\;2}/(2 {\mathcal{J}}_0) 
      + 
      h_{32} \, (\hat{T}_{3,+2} - \hat{T}_{3,-2}) \, ,
                                                                    \label{eq03} 
\end{equation}
where the first term which is proportional to the square of the nuclear 
angular momentum operator assures that the energy {\em vs.} spin dependence is
approximately quadratic. The second, tensor term, is responsible for the
tetrahedral symmetry of the hamiltonian (\ref{eq03}). The constant 
${\mathcal{J}}_0$ represents the effective moment of inertia. By construction, 
operators $\hat{T}_{3,\pm 2}$ are spherical tensors of rank $\lambda=3$ with 
the 'magnetic' components $\mu=\pm 2$; they are appropriately symmetrized 
functions of angular momentum operators  $\{ I_x, I_y, I_z \}$. The imaginary 
coefficient $h_{3,2}$ is a parameter of the model. According to the rotor 
formalism such a parameter must be independent of operators  $\{ I_x, I_y, I_z
\}$, but it may be a scalar function of the spin quantum  number $I$. Here we
use $h_{3,2}=c_{3,2}/I^2$, with an imaginary constant  $c_{3,2}$. This is a
convenient parametrisation that allows to keep the third order term dominated
by the usual, approximately quadratic energy dependence on spin. In the
following we use $c_{3,2}=i\,0.03$; it turns out that the exact value of this
constant has a very minor influence on the electro-magnetic transition rates,
the latter representing the main concern in this part of the discussion.

\begin{figure}[h]
\vspace{-0.15truecm}
\begin{center}
\includegraphics[height=7.0cm,width=8.5cm]{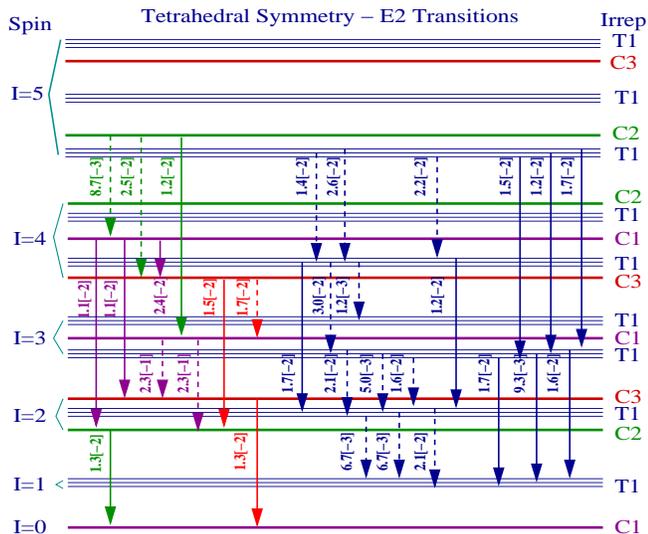}
\vspace*{-0.6truecm}
\end{center}\caption{  
         Color online) 
	 Tetrahedral rotor energy spectrum; the irreducible-representation 
	 symbols are marked on the right hand-side. Arrows indicate the strongest 
         E2-transitions [in units of $e^2(fm)^4$] depopulating two lowest states 
         of spin I=5, as an example. Full lines give stretched, dashed lines
         the non-stretched $\Delta I=1$ transitions. For easier legibility the 
         degenerate T1 states are 
         marked as split. Observe that the non-zero transitions connect only
         one-dimensional irrep states among themselves and the three-dimensional
         irrep states among themselves. In addition, there are no E2 transitions
         of the type $C1\to C1$, $C2\to C2$ nor $C3\to C3$.
         }
                                                                  \label{fig04}
\vspace*{-0.2truecm}
\end{figure}

The solutions of the Schr\"odinger equation with hamiltonian (\ref{eq03}) for
an even-even nucleus are characterized by integer spins and thus transform with
the help of the irreducible representations of the 'simple' $T_d$ group (as
opposed to the 'double' $T^D_d$-group of symmetry that applies for the nucleons
(spinors) mentioned earlier). The group $T_d$ has four irreducible representations,
three of them, denoted $C1$, $C2$ and $C3$ are one dimensional, the fourth one,
$T1$, is three-dimensional. Because of the latter, in the strict symmetry limit
the  rotational spectra must contain three-fold degenerate levels  (cf.
Fig.~\ref{fig04}). However, it is well known that quantum systems manifest 
zero-point motion oscillations: in the nuclear case the most important 
oscillations of this type are expected to be of quadrupole nature. We have
estimated, using collective model techniques with the Bohr Hamiltonian that the
expectation value of the quadrupole deformation in $^{110}$Zr is $\sim$ 0.05, 
depending somewhat on the details of the model. To simulate the zero-point 
motion effect in the following we therefore perform the calculations using a 
small quadrupole triaxial "contamination" term introducing three moments of 
inertia $J_x$, $J_y$ and $J_z$ which slightly deviate from $J_0$. These are 
calculated using uniform density ellispoidal shapes with quadrupole 
deformations $\beta=0.05$, $\gamma=30^{\text{o}}$ and $\alpha_{32}=0.15$. 

The zero-point motion mechanism simulated in this way has a very small
influence on the degeneracy (deviations of the order of 2 keV) but it has
a tremendous impact on the electro-magnetic transition probabilities. In
particular, at the exact symmetry limit the only transitions allowed are $E3$
ones; in the calculations with the zero-point motion effect included, the
quadrupole transitions are stronger by 5-6 orders of magnitude and the
non-stretched B(E2)s are usually slightly stronger than the stretched ones.

In the adiabatic approximation, the nuclear hamiltonian is a sum of the rotor
and the intrinsic energy term (cf. Eq.~(4-3) of Ref.\cite{ABo75}) and it
follows that the eigen-energies are also sums of two corresponding terms, while
the transition rates depend on the electric moments calculated with the help of
the intrinsic hamiltonian. The latter were calculated here using an uniform
charge distribution with the tetrahedral deformation of 0.15 and small
quadrupole deformations cited above. Results in Fig.~\ref{fig04} illustrate
only the rotor part of the spectrum. Consequently when reading them one has to
bear in mind that the energies of e.g. even-spin sequence and of the odd-spin
sequence may be shifted with respect to one another by an amount that needs to
be taken from the interpreted measured spectrum. The results of this kind could
also be used as a guidance for a band starting at $I=3$ (transitions
corresponding to $I\geq3$) or for the bands with all parities negative.

Moreover, it should be emphasized, that the reduced transition probabilities
for $(I\to I+1)$-transitions are usually some percentage stronger than those
for $(I\to I-1)$-transitions, so that if e.g. the odd-spin sequence is shifted
upwards, the inter-band transitions of the type $5\to6$, $4\to5$, etc. are to be
expected. In short: the information and prediction content of the diagrams
of the Fig.~\ref{fig04} type is much richer than what the first-glance reading
may suggest.

The partial transition pattern illustrated in Fig.~\ref{fig04} distinguishes
clearly between the decay modes of a tetrahedral quantum rotor and the one of
an ellipsoidal symmetry.  In the latter case, {\em only the stretched
E2-transitions are present}, connecting exclusively the band-members of the
same $D_2$-symmetric rotor irreducible-representations, as discussed in Chap.
IV of Ref.\cite{ABo75}. But even in the case of a strong triaxial minimum
competing ($\gamma\sim 30^o$) the reduced probabilities for the non-stretched
E2-transitions are nearly two orders of magnitude weaker compared to the
stretched ones. Moreover, they connect only the states {\em of the same
D$_2$-symmetry (same common irreps)} - while in the case of the T$_d$-symmetry,
the $(C_i\to C_i)$-transitions are forbidden, cf. caption to Fig.~\ref{fig04}.
These differences provide an excellent 'yes-no' criterion, provided that a
sufficient number of experimental transitions have been identified.

In the illustration of Fig.~\ref{fig04} we have selected a configuration
composed of the lowest lying rotational levels of the same parity with spins
$I\leq 5$; a more complete analysis will be presented elsewhere.

{\bf In summary:}   Low-lying tetrahedral configurations in Zr isotopes with 
$64\leq N \leq 72$ are obtained with both the HFB and Strutinsky techniques.  
The only noticeable difference between these very different realizations of  
the mean-field comes from the relative position of the tetrahedral minimum 
relative to the ground-state. HFB calculations with the SLy4 interaction 
predict a tetrahedral ground-state in $^{110-112}$Zr. The barrier heights 
which separate  the tetrahedral minima from their prolate counterparts have
also been  estimated and we emphasize that this region of the nuclear chart is 
potentially accessible for exploration using isomer and $\beta$-delayed
$\gamma$-ray spectroscopy following projectile fission.

The spectral properties expected to accompany the tetrahedral symmetry in
nuclei have been briefly discussed in terms of the electro-magnetic
transitions  in a quantum rotor with tetrahedral symmetry. The predicted decay
pattern is  very characteristic: it includes a competition between stretched
and  non-stretched E2-transitions and excludes the possibility of connecting
some  well defined groups of levels. This decay pattern is clearly distinct
from  known transition patterns of both the axial or nearly-axial rotors as
well as strongly tri-axial rotors whose spectra  contain principally the
E2-transition cascades. Since in the discussed nuclei  the tetrahedral-symmetry
minima are in competition with the ones of quadrupole type deformation, the
electro-magnetic decay criteria formulated offer a clear-cut distinction.

We believe that the predicted structure of the nuclear potential energy and the
corresponding minima strongly encourages the search for the first experimental
evidence of existence of these highly exotic nuclear symmetries.

{\bf Acknowledgements:} The authors are grateful to J. Dobaczewski for
letting them use the latest version of the {\sf HFODD} code prior to its
publication. This work is partially supported by EPSRC (UK), and the exchange
programme between IN2P3 (France) and Polish Nuclear Physics Laboratories.


\begin{thebibliography}{100}

\bibitem{JDu02}
         J. Dudek, A. G\'o\'zd\'z, N. Schunck and M. Mi\'skiewicz, 
         Phys. Rev. Lett. {\bf 88}, 252502 (2002)

\bibitem{TYM98}
         S. Takami, K. Yabana and M. Matsuo, 
         Phys. Lett. {\bf B431}, 242 (1998);
         M. Yamagami, K. Matsuyanagi and M. Matsuo, 
         Nucl. Phys. {\bf A693}, 579 (2001)
  
\bibitem{JDu02a}
         J. Dudek, A. G\'o\'zd\'z and N. Schunck, 
         Acta Phys. Polon. {\bf B34}, 2491 (2003) 
  
\bibitem{SMF87} 
         S. M. Fischer et al., Phys. Rev. Lett. {\bf 87}, 132501 (2001)

\bibitem{MBe94}
         M. Bernas et al., Phys. Lett. {\bf 331B}, 19 (1994)

\bibitem{ZPo00}
         Z. Podoly\'ak et al., Phys. Lett. {\bf 491B}, 225 (2000);\\  
         J. M. Daugas et al., Phys. Lett. {\bf 476B}, 213 (2000)

\bibitem{MNM01}
         M. N. Mineva et al., Eur. Phys. J. {\bf A11}, 8 (2001)

\bibitem{PFM03}
         P. F. Mantica et al., Phys. Rev. {\bf C67}, 014311 (2003);\\
         A. C. Morton et al., Phys. Lett. {\bf 544B}, 274 (2002);\\
         O. Sorlin et al., Eur. Phys. J. {\bf A16}, 55 (2003) 

\bibitem{JHH97}
         J. H. Hamilton et al., Prog. Part. Nucl. Phys. {\bf 38}, 273 (1997);\\ 
         A. G. Smith et al., Phys. Rev. Lett. {\bf 77}, 1711 (1996);\\
         M. A. C. Hotchkis et al., Nucl. Phys. {\bf A530}, 111 (1991) 
 
\bibitem{JFC84}
         J. F. Cornwell, 
        {\sl Group Theory in Physics}, (Academic Press, 1984)
  
\bibitem{GFK63}
         G. F. Koster, J. O. Dimmock, R. G. Wheeler and H. Statz, 
        {\sl Properties of the Thirty-Two Point Groups},
        (MIT Press, Cambridge, Massachusetts, 1963)
 
\bibitem{JDu87}
         J. Dudek, 
         Proc. Int. Winter Meeting on Nuclear Physics, Bormio, Italy 1987; 
        {\em (Ricerca Scientifica ed Educazione Permanente, Supplemento 
         N.56}, Edited by I. Iori).

\bibitem{TRW95}  
         T. R. Werner and J. Dudek, 
         Atomic Data Nucl. Data Tables {\bf 50}, 179 (1995)

\bibitem{KP002} 
         K. Pomorski and J. Dudek,
	        Phys. Rev. {\bf C67}, 044316 (2003)

\bibitem{VMS67} 
         V. M. Strutinsky, Nucl. Phys. {\bf A95}, 420, (1967);\\
         V. M. Strutinsky, Nucl. Phys. {\bf A122}, 1, (1968)

\bibitem{MdV83} 
         M. J. A. de Voigt, J. Dudek  and Z. Szyma\'nski, 
         Rev. Mod. Phys. {\bf 55} 949 (1983)

\bibitem{FRX02} 
         F. R. Xu, P. M. Walker and R. Wyss, 
         Phys. Rev. {\bf C65} 021303R (2002)

\bibitem{skalski}
         J. Skalski, S. Mizutori and W. Nazarewicz,
         Nucl. Phys. {\bf A617} 282-315 (1997)
  
\bibitem{PMo95}  
         P. M\"{o}ller, J. R. Nix, W. D. Myers and W. J. \'Swi\c{a}tecki, 
         Atomic Data Nucl. Data Tables {\bf 59}, 185 (1995)

\bibitem{WHP94}  
         W. H. Press, S. A. Teukolsky, W. T. Vetterling and B. P. Plannery, 
         {\em Numerical Recipes}, Cambridge University Press, 1994

\bibitem{Cha97} E. Chabanat, P. Bonche, P. Haensel, J. Meyer and R. Schaeffer,
	        Nucl. Phys. {\bf A635} 231 (1997)

\bibitem{hfodd} J. Dobaczewski and J. Dudek,
	        Comp. Phys. Comm.  {\bf 102}, 166 (1997); {\bf 102}, 183 (1997); 
		       {\bf 131} 164 (2000)\\
         J. Dobaczewski and P. Olbratowski, nucl-th/0401006
         {\em Submitted to Comp. Phys. Comm.}
\bibitem{Har93}
         W. G. Harter, {\sl Principles of Symmetry, Dynamics and Spectroscopy}
         (John Wiley $\&$ Sons, Inc., 1993)

\bibitem{ABo75}
         A. Bohr and B. R. Mottelson, 
         {\sl Nuclear Structure, Vol II}, (W. A. Benjamin, Inc., 1975)
  
\vspace*{-0.1cm}

\end{thebibliography}
\end{document}